# Large Thermal Hall Effect in $MnPS_3$

**Mohamed Nawwar[1], Robin R. Neumann[3,4], Jiamin Wen[1], Ingrid Mertig[3], Alexander Mook[4, *, †], and Joseph P. Heremans[1,2,5, *, ‡]**

[1] Department of Materials Science and Engineering, The Ohio State University, Columbus OH, USA
[2] Department of Mechanical and Aerospace Engineering, The Ohio State University, Columbus OH, USA
[3] Institut für Physik, Martin-Luther-Universität Halle-Wittenberg, D-06099 Halle (Saale), Germany
[4] Institut für Physik, Johannes Gutenberg Universität Mainz, D-55128 Mainz, Germany
[5] Department of Physics, The Ohio State University, Columbus OH, USA

* Corresponding author
† Contact author: amook@uni-mainz.de
‡ Contact author: Heremans.1@osu.edu

E-mail: xxx@xxx.xx

## Abstract

Recent studies have demonstrated that the thermal Hall effect can originate from magnons (magnon Hall effect), phonons (phonon Hall effect), or their combination (magnon-polaron Hall effect). The magnon-polaron Hall effect, first observed in $Fe_2Mo_3O_8$, is particularly intriguing as its thermal Hall signal can be remarkably large. In this study, we explore the thermal Hall effect in $MnPS_3$, an insulating antiferromagnetic material exhibiting a spin-flop transition and significant magnetoelastic coupling, making it a strong candidate for studying the thermal Hall effect originating from spin-lattice coupling. We report an exceptionally large thermal Hall angle down to 4 K and show that it cannot be accounted for by standard calculations based on the intrinsic magnon-polaron Berry curvature. Our findings provide an in-depth analysis of the role of the spin-flop transition in the thermal properties of $MnPS_3$ and call for further theory development on magnon-phonon coupling and scattering to reveal their influence on transverse heat transport.

Keywords: term, term, term

---

## 1. Introduction

The search of materials hosting charge-neutral topological excitations has gained significant interest in recent years [1,2]. Experimental investigations of electronic topological excitations are facilitated by high-sensitivity techniques like voltage measurements. Measurements such as the electrical Hall effect and angle-resolved photoemission spectroscopy (ARPES) have been instrumental in confirming electronic topological states [3]. However, detecting charge-neutral topological quasiparticles like magnons has proven far more challenging, as their detection relies on techniques such as thermal Hall effect (THE) measurements and inelastic neutron scattering (INS) [4]. Early studies of THE in ferromagnetic insulators suggested the presence of topological magnons and attributed the non-zero Berry curvature responsible for these effects to the Dzyaloshinskii-Moriya interaction (DMI) [5–7].

Subsequent research revealed that magnetic long-range order is not necessary for the emergence of THE, indicating that magnetic quasiparticles are not the only source of transverse heat transport. Evidence for phonons contributing to THE grew stronger as THE was observed in the paramagnetic regime of insulating magnets and in non-magnetic insulators [8–11]. In addition to THE, the coupling between magnons and phonons has been shown to impact the spin Seebeck effect in YIG [12]. These findings showed that the role of phonons cannot be ignored when analyzing THE in insulating magnets and that the interplay between magnons and phonons is crucial.

Studying materials with strong spin-lattice coupling was essential to understand the topological nature of this interplay. Thermal Hall measurements on $Fe_2Mo_3O_8$, a multiferroic material with a strong spin-lattice coupling, demonstrated a record-high THE at the time [13]. Detailed INS studies confirmed the anti-crossing between phonon and magnon bands, providing experimental evidence of the topological gap [14]. The resulting non-zero Berry curvature from this hybridization mechanism was proposed to contribute to the observed THE [15].

$MnPS_3$, a two-dimensional insulating magnet, emerges as a promising candidate for exploring magnon-phonon hybridization due to its significant magnetoelastic coupling and Néel AFM order [16], which can be manipulated with magnetic fields. $MnPS_3$ has been shown to host the spin Nernst effect, which has been attributed to the non-zero Berry curvature of the magnon bands [17–19]. Moreover, it has been theoretically proposed that the coupling between its magnon and phonon bands creates a topologically non-trivial gap, resulting in a finite THE [20]. The material also undergoes a spin-flop transition, altering its spin structure and magnon bands [21]. Computational studies have shown the effect of the spin-flop transition on the expected THE [22,23]. A spin-flop transition potentially alters the magnon-phonon hybridization in the material and may have a strong impact on the THE. $NiPS_3$, a sister compound to $MnPS_3$, was shown to host a large THE (~2 W $m^{-1}$ $K^{-1}$), but no spin-flop transition was observed up to 50 T. The authors attributed the large THE to magnon-phonon hybridization [24]. These properties make $MnPS_3$ a compelling candidate to further study thermal Hall effects induced by magnon-phonon interactions.

In this letter, we report our investigation of thermal transport properties in $MnPS_3$. We measured thermal conductivity ($\kappa_{xx}$), thermal Hall conductivity ($\kappa_{xy}$), and heat capacity as functions of magnetic field across a range of temperatures, down to 4 K. The thermal transport data exhibits distinct behavior directly linked to the spin-flop transition in $MnPS_3$. Notably, the thermal Hall conductivity ($\kappa_{xy}$) demonstrates a non-monotonic dependence on the magnetic field and undergoes a sign change around 10 K, reaching an astounding value of -14 W $m^{-1}$ $K^{-1}$. We present multiple observations that point toward significant magnon-phonon interactions, but a magnon-phonon hybridization model cannot account for the magnitude of the observed effect. This observation suggests that instead of the intrinsic magnon-polaron Berry curvature, magnon-phonon scattering might be more relevant for the explanation of the measurements. We discuss experimental and theoretical challenges with respect to $MnPS_3$ and thermal Hall measurements more generally.

## 2. Results and Discussion

### *2.1 Crystal Structure and Magnetization*

The crystal structure of $MnPS_3$ belongs to the monoclinic space group *C2/m* (**Fig. 1(a)**) [25]. It is an insulating antiferromagnet with a Néel temperature of 78 K and an electronic band gap of 2.79 eV [26,27]. $Mn^{2+}$ ions are arranged on a honeycomb lattice in a Néel antiferromagnetic configuration, where spins are oriented along the out-of-plane direction (**Fig. 1(a)**) [21]. The spins exhibit a canting angle of approximately 8° toward the *ab*-plane [28].

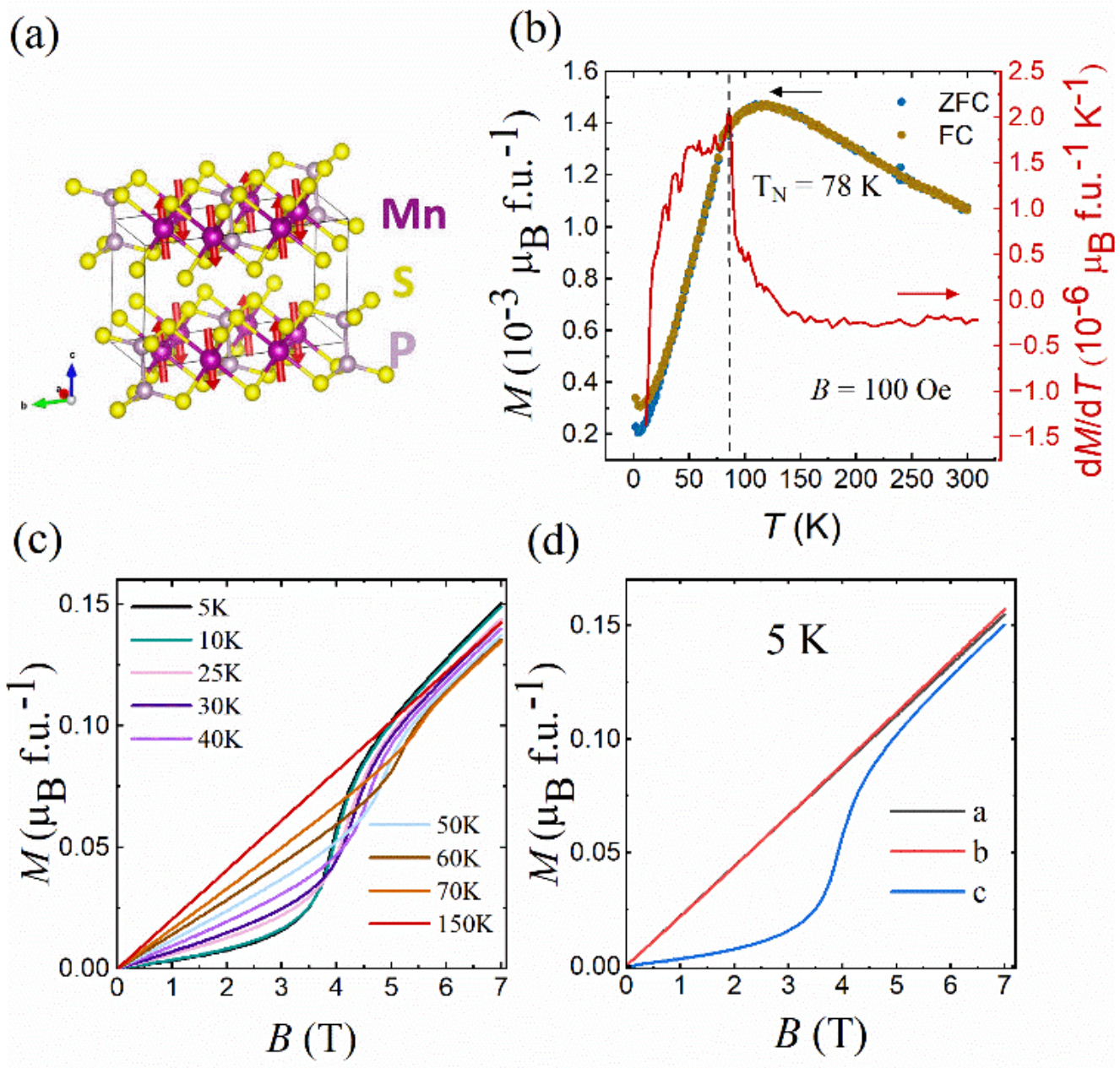

**Figure 1**. Crystal structure and magnetic properties of $MnPS_3$. (a) Monoclinic crystal structure of $MnPS_3$. $Mn^{2+}$ ions are arranged in a honeycomb lattice with spins pointing along the c-axis with a slight tilt. (b) Magnetization data as a function of temperature shows the magnetic transition temperature at 78 K. The Néel temperature was extracted from the maximum point of the derivative dM/dT. (c) Magnetization as a function of magnetic field (applied along the c-axis) at different temperatures showing the spin-flop transition. (d) Magnetization vs external magnetic field applied along three different axes (a, b and c) at 5 K. No anisotropy between a and b axes was observed.

The Néel temperature was determined by identifying the peak in the derivative of magnetization with respect to temperature (*dM/dT*), yielding a value of 78 K, consistent with previous findings (**Fig. 1(b)**) [21]. A small separation between the zero-field cooled (ZFC) and field cooled (FC) curves was observed at low temperatures ~20 K. This behavior is attributed to spin canting in $Mn^{2+}$ ions, which can induce a weak ferromagnetic moment at a low magnetic field [21], a phenomenon further confirmed in our magnetization measurements (**Fig. S1**). Additionally, no significant difference was observed between the magnetization along the *a* and *b* axes (**Fig. 1(d)**).

Upon applying a strong magnetic field along the *c* axis, the sample undergoes a broad spin-flop (SF) transition (**Fig. 1(c)**). The sharpness of the transition increases with lower temperatures. At 5 K, the spin-flop transition is the sharpest, beginning at 3.5 T, where the magnetization deviates from its linear behavior, and ending at 5 T, where the magnetization re-enters the linear regime. The magnetic transition region broadens, and the SF field increases with increasing temperature, as clearly illustrated in **Fig. S1(d)**.

## *2.2 Thermal Conductivity*

The thermal conductivity ($\kappa_{xx}$) of $MnPS_3$ (sample 1) at 0 T (**Fig. 2(b)**) is consistent with a previous report [29], reaching a peak value of 290 W m$^{-1}$ K$^{-1}$ at 14 K. In the low-temperature regime (<14 K), $\kappa_{xx}$ follows a $T^{1.5}$ dependence. Above 14 K, the thermal conductivity decreases due to Umklapp scattering. Thermal transport data was consistently reproduced across four different samples, each using varied thermometry methods (**Fig. S4**). Unlike thermal conductivity, the heat capacity ($C_v$) in $MnPS_3$ follows the conventional $T^3$ dependence at low temperatures as shown in **Fig. 2(c)**.

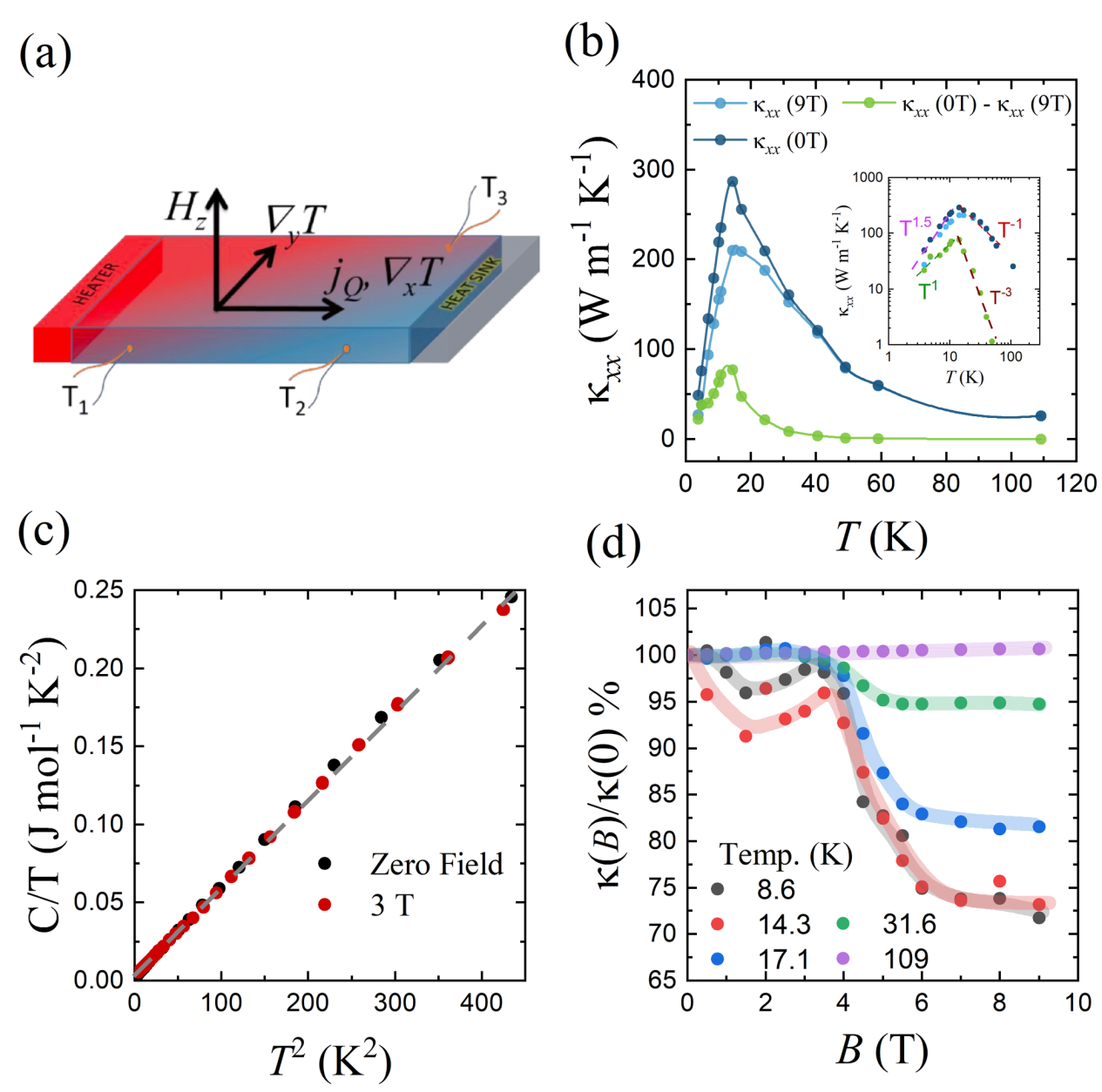

**Figure 2** Thermal conductivity and heat capacity of $MnPS_3$. (a) Schematic of the thermal transport measurement setup. (b) Thermal conductivity $\kappa_{xx}$ ($B_z$) for sample 1 as a function of temperature down to 4 K. Inset shows the log-log plot of the thermal conductivity (c) Heat capacity at low temperatures showing a $T^3$ behavior at both zero applied magnetic field and 3 T. (d) Thermal conductivity behavior as a function of magnetic field at fixed temperatures.

**Figure 2(d)** highlights the magnetic field dependence of thermal conductivity. Below the Néel temperature (78 K), $\kappa_{xx}$ ($B_z$) exhibits a distinct trend: it remains nearly constant until approximately 4 T, then sharply decreases and stabilizes again at a lower value. This behavior is strongly correlated with the spin-flop transition field. Above the Néel temperature, $\kappa_{xx}$ ($B_z$) is nearly constant. The abrupt drop in thermal conductivity and the temperature-dependence indicate an enhancement in the magnon-phonon scattering at the spin-flop transition. A similar behavior has been previously observed in other materials such as $Ni_3TeO_6$, $Fe_2Mo_3O_8$, and $MnBi_2Te_4$ [13,30,31].

## 2.3 Thermal Hall Effect

**Figures 3(a, b)** shows thermal Hall angle (tan θ = $\kappa_{xy}$ / $\kappa_{xx}$) and thermal Hall conductivity, $\kappa_{xy}$($B_z$), as a function of magnetic field at five different temperatures. At 109 K, which lies in the paramagnetic regime above the Néel temperature, the thermal Hall conductivity signal is negligible, indicating that magnetic ordering plays a crucial role in inducing the thermal Hall effect. At temperatures below the Néel temperature, the observed thermal Hall signal exhibits a non-monotonic dependence on the magnetic field, characterized by an increase up to a maximum value, followed by a decrease, forming a pronounced peak. This behavior aligns closely with the spin-flop transition observed in the sample. Due to this non-monotonic nature of $\kappa_{xy}$ ($B_z$), the peak value was extracted manually and used to analyze its temperature dependence (**Fig. 3(d)**). $\kappa_{xy}$ ($B_z$) data for all temperatures and samples are provided in supplementary materials.

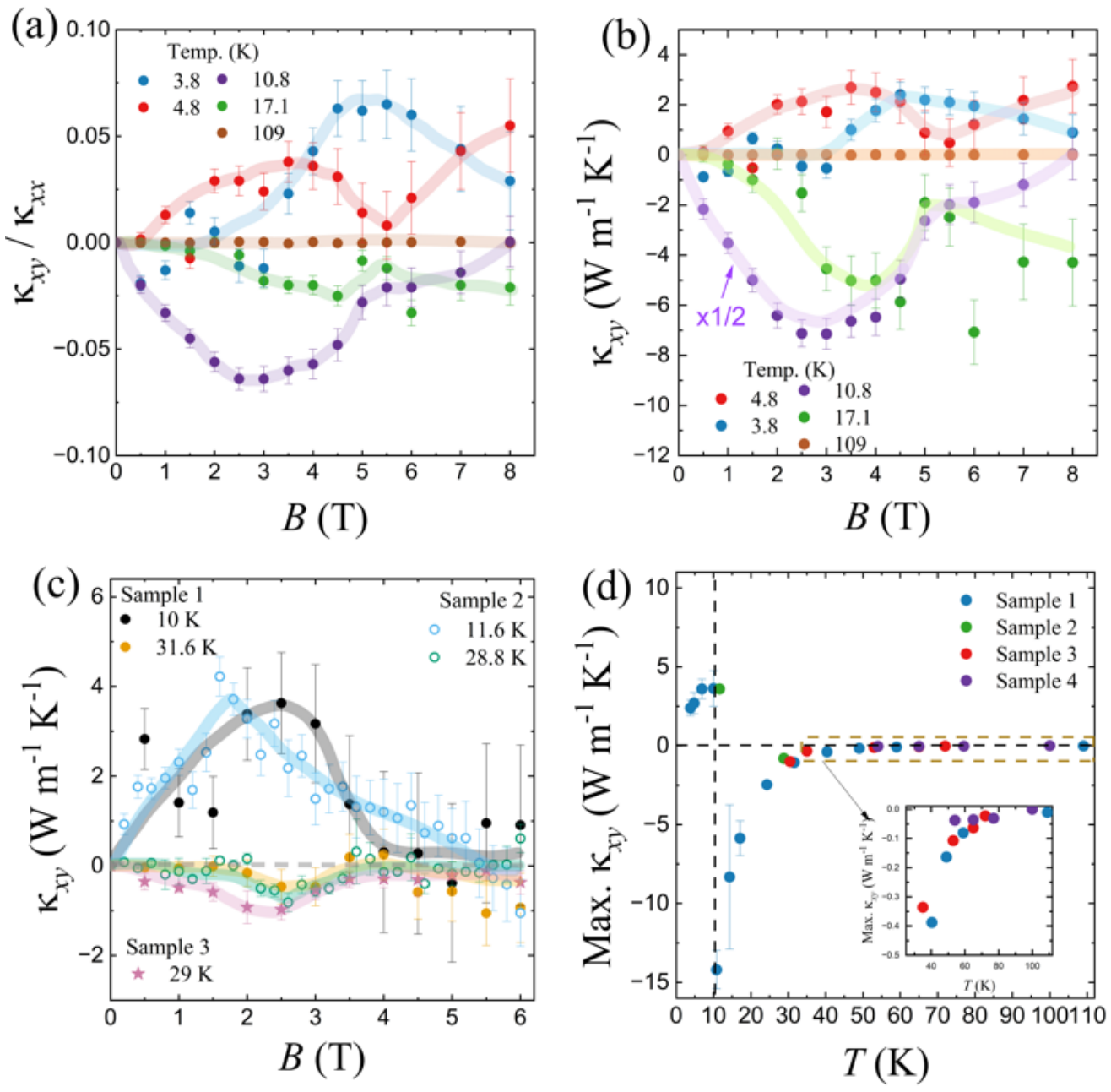

**Figure 3** Experimentally observed thermal Hall conductivity ($\kappa_{xy}$ ($B_z$)) of $MnPS_3$. (a) Thermal Hall angle (*tan θ = $\kappa_{xy}$ / $\kappa_{xx}$*) of sample 1 as a function of externally applied magnetic field along the c-axis at selected temperatures. The field dependent *thermal Hall angle* exhibits a sharp non-monotonic behavior consistent with the spin-flop transition. (b) $\kappa_{xy}$ of sample 1 exhibiting similar behavior to *tan θ*. (c) $\kappa_{xy}$ at two fixed temperatures around 10 K and 30 K across three different samples reproducing the sign-change and the non-monotonic behavior. (d) Maximum $\kappa_{xy}$ as a function of temperature. The maximum values were manually extracted from the field-dependent data. A sharp sign-change was observed around 10 K.

The temperature dependence of $\kappa_{xy}$ exhibits two distinct regimes with opposite signs: a low-temperature regime (<10 K) exhibiting a positive sign and a high-temperature regime (>10 K) with a negative sign. This behavior was successfully reproduced across multiple samples (**Fig. 3(c))**This temperature-induced sign reversal coincides with the peak in magnon-phonon scattering, as reflected in the difference between zero-field and high-field thermal conductivity ($\kappa_{xx}$ (0 T) - $\kappa_{xx}$ (9 T)). The magnitude of these effects suggests that magnon-phonon interactions play an essential role in the thermal Hall effect (**Fig. S5**).

## 2.4 Spin Flop

**Figure 4(a)** shows the magnetization curve at 5 K, revealing that the spin-flop transition field begins around 3.5 T and ends around 5 T. Above 5 T, the sample enters a state where the spins are forced to align increasingly along the external magnetic field; however, full saturation is not observed up to 7 T. This behavior is reflected in the heat capacity curve (**Fig. 4(b)**) which increases linearly up to about 4 T, where the system enters the spin-flop transition. Between 4 T and 5 T, the heat capacity exhibits a slight decrease, followed by a more pronounced drop above the completion of the spin-flop transition. Heat capacity conveys a thermal picture of the occupied magnon and phonon density of states. Below the spin-flop transition, the magnetic field shifts the two magnon branches upwards and downwards due to Zeeman energy (**Fig. 4(e)**). According to Bose-Einstein statistics, the thermal population in the upper branch decreases marginally, while the lower branch occupation increases exponentially, resulting in an overall rise in heat capacity. At the spin-flop transition, the lower magnon branch becomes a zero-energy Goldstone mode (**Fig. 4(f)**), leading to maximum thermal occupation and, consequently, maximum heat capacity. Above the spin-flop transition, the upper branch shifts further upward, reducing thermal occupation and causing a decline in heat capacity.

Thermal conductivity ($\kappa_{xx}$) and the thermal Hall effect ($\kappa_{xy}$) exhibit a strong correlation with the spin-flop transition as well (**Figs. 4(c, d)**). The interplay between the magnon and phonon modes provides a plausible explanation for both behaviors. In the case of thermal conductivity $\kappa_{xx}$, the primary contributions are specific heat capacity and the scattering of the heat carriers. However, $\kappa_{xx}$ exhibits a distinct and more pronounced behavior compared to specific heat. While specific heat varies by only ~4%, thermal conductivity remains nearly constant at low fields and drops sharply by approximately 40% around the spin-flop transition. The thermal conductivity is directly related to heat capacity ($C$) through the relationship, $\kappa = \frac{1}{3} C v^2 \tau$, where $v$ is the sound velocity and $\tau$ is the relaxation time. Since sound velocity is constant with the applied magnetic field, and $C$ varies only slightly, we conclude that magnon-phonon scattering ($\tau$) is the dominant term in $\kappa_{xx}$ ($B_z$).

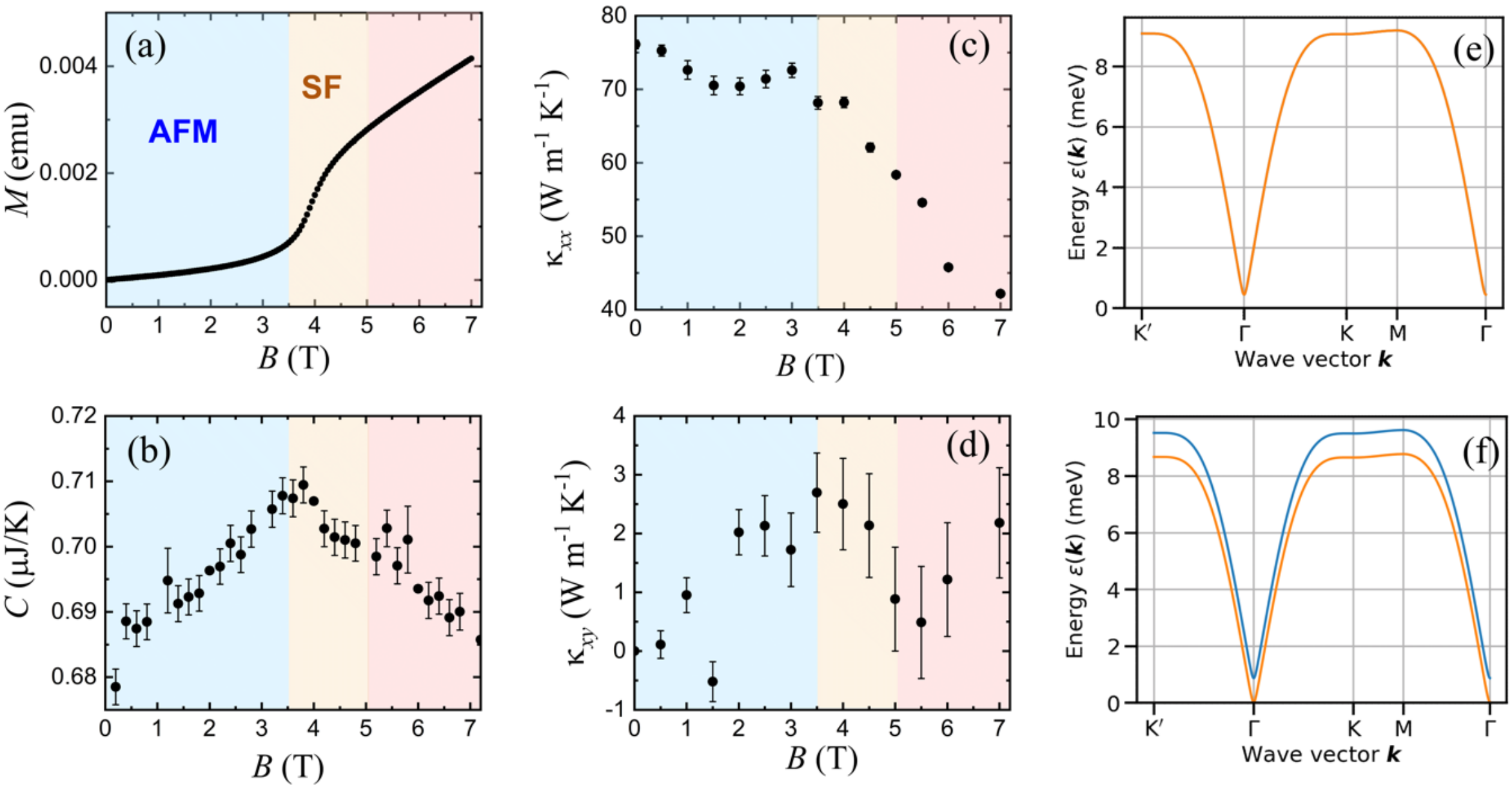

**Figure 4** Impact of spin-flop (SF) transition on the thermal properties of $MnPS_3$ at 5 K. (a) Magnetization data as a function of externally applied magnetic field along the c-axis shows clear signatures of the SF transition starting around 3.5 T at 5 K. (b) Heat capacity at 5 K as a function of magnetic field shows strong dependence on the magnetic phase transitions as it increases in the AFM phase and then sharply decreases when it enters the SF phase. (c, d) Magnetic field dependence on thermal conductivity and thermal Hall conductivity at 5 K, respectively. (e, f) Calculations of the magnon band structure at 0.01 T and 4 T, respectively. Blue and orange bands represent the polarization (spin up and down).

These observations suggest that magnon-phonon scattering plays a vital role in thermal conductivity, as the field-induced reduction of the magnon energies provides more efficient scattering channels during the spin-flop transition, thereby enhancing scattering.

## *2.5 Magnon-Phonon Hybridization Model*

To explore the origin of the observed thermal Hall signal and its non-monotonic behavior, we calculated the intrinsic Berry curvature arising from the hybridization of magnon and phonon bands that leads to hybrid quasiparticles known as magnon polarons [20,32–36]. We modeled a monolayer of $MnPS_3$ by a Hamiltonian that incorporates magnetic, elastic, and magnetoelastic components. The magnetic component of the Hamiltonian includes the Heisenberg exchange parameters ($J_1$, $J_2$, $J_3$), the Dzyaloshinskii-Moriya interaction (DMI; $D$), an eays-axis anisotropy ($K$), and a Zeeman term (with Lande´ factor $g$). The lattice component is represented by a spring model for the out-of-plane displacements of the Mn ions between nearest neighbors (elastic constant $C$). The magnetoelastic component couples the spins to the out-of-plane displacements (magnetoelastic constant $\lambda$).

The parameters used for this model were obtained from ab-initio calculations [16], except for the elastic constant, which was obtained from a separate experiment that we conducted using resonant ultrasound spectroscopy. For calculating the thermal Hall conductivity, we exclusively considered the Berry curvature-driven intrinsic contribution and neglected anharmonicities such as magnon-magnon interactions and any additional scattering effects. Further details are given in **Methods**.

**Figures 5(a)** and **5(b)** depict the results of the calculated $\kappa_{xy}$ obtained from the magnon-phonon hybridization model. The magnetic field dependence of $\kappa_{xy}$ features a nonmonotonic behavior (**Fig. 5(a)**). At lower temperatures ($T$ = 1 and 2 K), $\kappa_{xy}$ exhibits an increase with the applied magnetic field, reaching a peak value below the SF transition, which occurs at 4.26 T for the chosen parameters. Beyond this transition, $\kappa_{xy}$ is strongly suppressed. As temperature rises, the sign of $\kappa_{xy}$ is reversed from positive at below 3 K to negative above at 3 K.

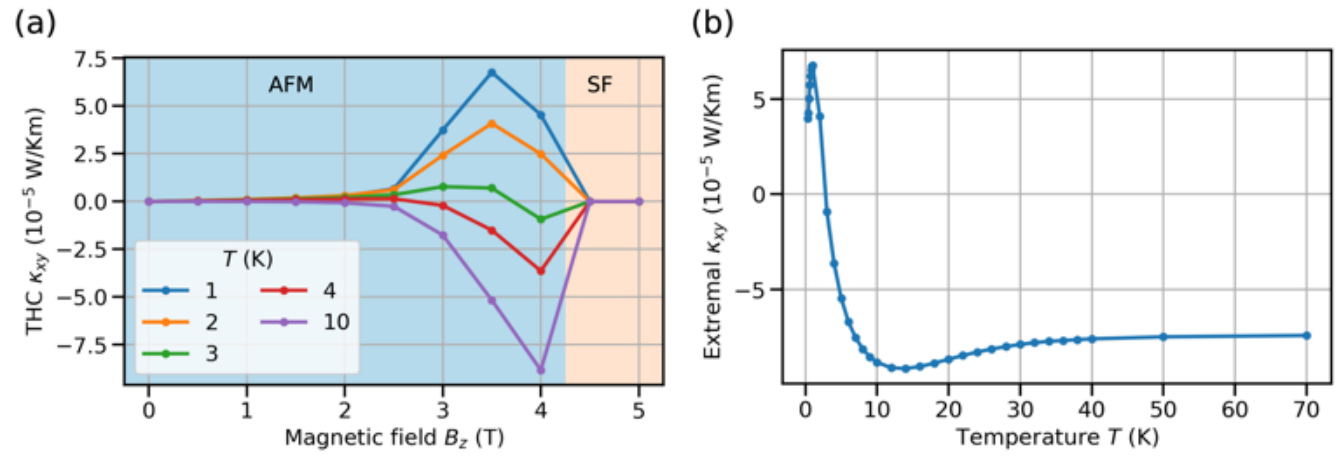

**Figure 5** Calculations of thermal Hall conductivity based on the magnon-polaron intrinsic Berry curvature. (a) Calculated $\kappa_{xy}$ as a function of an externally applied magnetic field. A non-monotonic behavior is observed and attributed to the spin-flop transition. (b) The temperature dependence of the peak $\kappa_{xy}$ exhibits a sign change.

The effect of temperature on $\kappa_{xy}$ is further illustrated in **Fig. 5(b)**, where the peak values of the $\kappa_{xy}$($B_z$) curves are plotted as a function of temperature, similar to our treatment of the experimental data. $\kappa_{xy}$(T) changes sign near 3 K, reaches a minimum at 12 K, and eventually saturates. Note that at higher temperatures (relative to $T_N$), magnon-magnon interactions are expected to suppress the intrinsic thermal Hall. Our model based on linear spin-wave theory is not reliable in this limit.

## *2.6 Berry Curvature Evolution with Magnetic Field*

To scrutinize the behavior of $\kappa_{xy}$ and its correlation with the spin-flop transition, we calculated the evolution of the magnon band structure and the Berry curvature of the lowest band with the magnetic field as shown in **Fig 6**. At zero field and without magnetoelastic coupling, the spin-up and spin-down magnon bands are degenerate. By introducing magnetoelastic coupling, the degeneracy is lifted, resulting in hybridization between magnon and phonon bands (**Fig. 6(a)**);

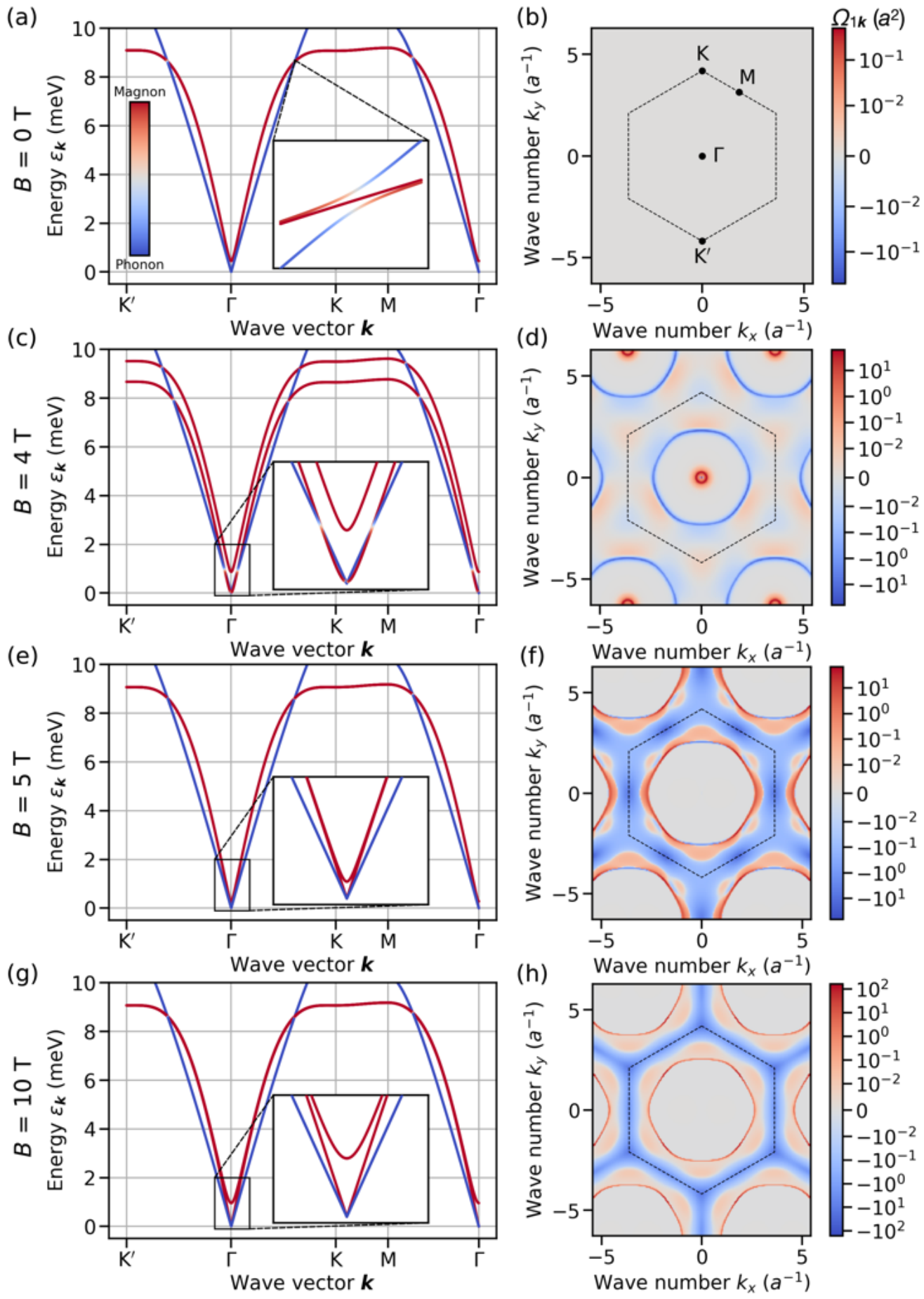

**Figure 6** Model calculations of the magnon and phonon dispersion relations and the corresponding Berry curvature at selected values of magnetic fields. (a) Band structure along a high-symmetry path at $B_z = 0$ T. Inset: Zoom into the avoided crossing between two magnons and one phonon band. The optical phonon band has been cut off for better visibility. Red/blue/gray colors indicate magnon-like/phonon-like/hybrid quasiparticle states. (b) Berry curvature of the lowest band in reciprocal space. The dashed hexagon demarcates the first Brillouin zone. Red/blue/gray colors indicate positive/negative/zero Berry curvature, respectively. (c, d), (e, f) and (g, h) same as (a, b) for B = 4, 5, and 10 T, respectively.

however, the bands are no longer spin-polarized. Furthermore, the Berry curvature of the lowest band is zero (**Fig. 6(b)**). By increasing the magnetic field, the two magnon bands are split due to the Zeeman interaction, with one band shifted upwards and one downwards. The downward shift of the lower magnon band results in a second avoided crossing at low energies with the acoustic phonon band (**Fig. 6(c)**). These two avoided crossings create pronounced Berry curvature rings centered around the Γ point with different radii and opposite signs (**Fig. 6(d)**). At $B_z = 4.26$ T, the lower magnon band reaches zero energy, and the antiferromagnetic phase becomes unstable and transitions to the spin-flop phase. Consequently, the Néel vector is rotated from its out-of-plane to an in-plane orientation, and a ferromagnetic moment is developed along the externally applied magnetic field. As an effect of the spin-flop transition, the magnon bands approach each other (**Fig. 6(e)**). The lower magnon band forms a Goldstone mode with linear dispersion, whose velocity exceeds that of the transverse phonons, resulting in the disappearance of the low-energy avoided crossing (**Fig. 6(f)**). This eliminates the low-energy Berry curvature ring observed in the antiferromagnetic phase and explains the suppression of $\kappa_{xy}$ in the spin-flop phase as observed in **Fig. 5(a).**

As the ferromagnetic moment increases with the applied magnetic field, the upper magnon band gradually changes its curvature, while the lower one is pinned at zero energy due to the isotropy of the ground state energy with respect to the in-plane Néel vector, preserving the Goldstone mode (**Fig. 6(g)**). Because the bands are mostly modified at the Γ point, the location of the high-energy avoided crossing remains constant over a large range of magnetic field (**Fig. 6(h)**). Hence, $\kappa_{xy}$ does not strongly depend on $B_z$ above the spin-flop phase.

## 3. Discussion

Although the experimental and the calculated $\kappa_{xy}$ share some qualitative similarities, there is a large difference in magnitude. The discrepancy can be related to complications in the experimental measurements as well as assumptions of the model. The measurement of thermal Hall conductivity, which corresponds to the odd part of the thermal conductivity tensor, is challenging because it relies on several assumptions. It is commonly known that the transverse thermocouples can pick up signals from the longitudinal thermal resistivity due to small misalignments. These can be eliminated by measuring the transverse temperature gradient at positive and negative magnetic fields and antisymmetrize them because only the thermal Hall conductivity is odd under magnetic field, according to the Onsager relation [37]. This, however, assumes that an applied magnetic field also reverses the microscopic magnetic structure. This assumption is violated in materials with a magnetic hysteresis in *M(B)*. Although our data suggests a finite ferromagnetic component, we have verified that the hysteresis does not persist above 0.3 T (**Fig. S1**), which is below the smallest measured magnetic field data point in the thermal transport experiment. Nevertheless, we cannot exclude that the magnetic components hidden in *M(B)* do not follow the magnetic field, which is a potential issue for antiferromagnets in general.

Another complication arises from the anisotropy of the material. Because of the monoclinic structure and the tilting of the Néel vector in $MnPS_3$, $\kappa_{xx}$ and $\kappa_{yy}$ do not have to be equal. In the experiment, the thermal resistivity $w_{xx}$ is measured. In order to convert to conductivities, both $w_{xx}$ and $w_{yy}$ would be required, which cannot be obtained in a single measurement geometry. Hence, we had to assume $w_{xx} = w_{yy}$.

On the side of the model, we have focused on one particular mechanism, which originates from the intrinsic Berry curvature of magnon-phonon hybrids. For the magnetoelastic coupling, we have focused on a particular interaction derived from spin-orbit coupling that exclusively hybridizes magnons and out-of-plane phonons [16,20,32,38–40]. In the spin-flop phase, other symmetry-allowed interactions can also lead to hybridizations, including in-plane phonons [23,41]. However, these would introduce more parameters which have not been calculated for this material yet. They could give rise to a non-zero $\kappa_{xy}$ in the spin-flop phase if the longitudinal acoustic phonon bands cross with the magnon bands. Within the antiferromagnetic phase, the considered form of magnetoelastic coupling is dominant [32], which is why we do not expect other forms of magnetoelastic couplings to strongly modify our results below ~4 T.

Another source of transverse currents due to time-reversal symmetry breaking may arise from many-body magnon-magnon, phonon-phonon, and magnon-phonon interactions, which not only can renormalize the Berry-curvature driven contribution studied here [42–47], but can give rise to skew scattering and side jump as an alternative mechanism [48–51]. Scattering off defects can further contribute to the thermal Hall effect [52]. These theories exceed the scope of this project.

Considering the model, it should be noted that experimental findings that challenge the out-of-plane Néel order in $MnPS_3$ have started a debate on the magnetic structure. Our model does not reproduce the complex temperature-dependent magnetic phase diagram reported here and elsewhere [21]. In particular, the ferromagnetic moment below 20 K cannot be accounted for. This should be addressed in the future by constructing a magnetic Hamiltonian capturing the magnetic ground states.

While the observed values of $\kappa_{xy}$ in our study are relatively high, it is worth noting that comparable magnitudes have been reported in another material of the $MPX_3$ system, $NiPS_3$ [24]. The authors of the study reported values of $\kappa_{xx}$ up to 300 W/mK and $\kappa_{xy}$ reaching 2 W/mK, which are of the same order as those observed in our measurements. Notably, $NiPS_3$ does not undergo a spin-flop transition up to 50 T and exhibits a different spin structure compared to $MnPS_3$. Consequently, $\kappa_{xy}$ in $NiPS_3$ displays a linear field dependence, in contrast to the non-monotonic behavior observed in our system. Additionally, $\kappa_{xx}$ did not exhibit any major reduction up to 14 T, which could also be attributed to the lack of spin-flop transition. The authors of that study also highlighted the role of crystal twinning in affecting the magnitude of $\kappa_{xy}$ and the magnetic anisotropy. However, in our crystals, we did not observe a significant reduction of $\kappa_{xy}$ across different crystals, and optical microscopy inspection revealed no prominent twinning features similar to those reported in $NiPS_3$. Therefore, we consider the effect of crystal twinning to be negligible in our samples and have not included it as a contributing factor in our analysis.

During the preparation of this manuscript, we became aware of a similar study on $MnPS_3$ that reached a different conclusion [41]. Our sample exhibited $\kappa_{xy}$ values that are two orders of magnitude larger and displayed a pronounced non-monotonic behavior. In contrast, the other study reported a monotonic $\kappa_{xy}$ with a small hump near the spin-flop transition. We suspect that this discrepancy could stem from sample quality as our thermal conductivity data is three times higher than the one they reported. We also note that our thermal conductivity values agree with a separate study [29].

## 4. Conclusion

In this study, we investigated the thermal transport properties of $MnPS_3$, focusing on the magnetic field dependence of thermal conductivity ($\kappa_{xx}$) and thermal Hall conductivity ($\kappa_{xy}$). The observed behavior of $\kappa_{xx}$ points to significant magnon-phonon scattering above the spin-flop transition. The thermal Hall conductivity ($\kappa_{xy}$) exhibited a non-monotonic dependence on the magnetic field.

Within our model, the anti-crossing between magnon and phonon bands is maximized below the spin-flop transition, generating a non-zero Berry curvature near the Γ-point. As the applied magnetic field increases, the slope of the lower magnon band steepens, causing the hybridization point and the associated Berry curvature to vanish, thus generating a non-monotonic $\kappa_{xy}$. Furthermore, the model produces a low-temperature sign reversal in $\kappa_{xy}$ through the calculated intrinsic Berry curvature, which features contributions of opposite signs around the Γ-point.

While our magnon-phonon hybridization model successfully reproduces some qualitative features of the observed thermal Hall effect (THE), it does not account for the exceptionally large magnitude of $\kappa_{xy}$. Therefore, our findings call for a better understanding of the magnetic structure of $MnPS_3$ and the exploration of alternative mechanisms, such as magnon-phonon skew scattering, that give rise to a thermal Hall effect.

## 5. Methods

### *5.1 Synthesis and material characterization*

$MnPS_3$ single crystals were grown using Chemical Vapor Transport (CVT) technique. Stoichiometric amounts of elemental powders and 0.03 g of iodine (transport gas) were mixed together in an Ar-filled glovebox and then loaded into a fused silica tube and sealed under vacuum (<1 Torr). During the sealing, the precursors were immersed in liquid nitrogen to prevent the evaporation of iodine. The sample was then placed in a two-zone furnace with $T_{hot}$ at 680 °C and $T_{cold}$ at 650 °C for one week with a ramp rate of 5 °/min and cooling rate of 0.1 °/min. The resulting crystals were green and shown in **Figure S2**. Sample purity was confirmed using x-ray diffraction (XRD) (**Fig. S2).** Magnetization measurements were conducted using a Quantum Design Magnetic Property Measurement System (MPMS) with maximum field of 7 T. A single crystal was mounted on a quartz tube using GE varnish and measured in DC mode. Heat capacity was conducted on the Quantum Design Physical Property Measurement System (PPMS) with a maximum field of 9 T using the heat capacity mode and puck; the sample was mounted with the magnetic field pointed along the c-axis.

### *5.2 Thermal Transport*

Thermal transport measurements were conducted using Quantum Design PPMS. The thin fragile sample was mounted on our home-made setup for thermal transport (**Fig. S3**). The sample was mounted such that the magnetic field is pointing out-of-plane, along the *c*-axis, and the heat flux is sent across the *ab*-plane. The whole mount was made from Kapton tape, which has a very small thermal conductivity so that it would be shorted out by the high thermal conductivity of the sample. We used GE varnish to connect the sample to the heat sink and the heater. Thermocouples were connected to the sample using silver epoxy.

We used a 120 Ω resistor to send heat along the *x*-axis and a heat sink made from LiF. The measurements were conducted in a steady-state mode, where we waited for over 10 min after the heater was turned on to ensure thermal equilibrium across the sample. Additionally, we waited for five minutes after the magnetic field was applied at each step to avoid any extrinsic magnetocaloric effects. All voltage measurements across thermocouples were conducted using a 2182A Keithley nanovoltmeter. At low temperatures, we averaged the measured voltage for a few minutes (5 - 10 min) to lower the error as much as possible.

We used different thermometry across the samples. In samples 1 and 2, we used type E thermocouples, while with samples 3 and 4, type T thermocouples were used. Sample 1, three thermocouples were connected differentially, while a separate fourth thermocouple was mounted on the sample to measure sample temperature. Samples 2, 3 and 4 used three separate thermocouples. The longitudinal ($\Delta T_x$) and transverse ($\Delta T_y$) temperature differences were calculated by directly converting the measured voltages using the thermocouples standardized Seebeck polynomials. $\Delta T_y$ was then anitsymmetrized to remove the contact misalignment. $\Delta T_x$ and $\Delta T_y$ were then used to calculate $\kappa_{xx}$ and $\kappa_{xy}$ using Fourier's law of heat conduction.

## 5.3 Magnon-phonon hybridization model

*5.3.1 The model.* We describe a monolayer of $MnPS_3$ by the Hamiltonian $\mathcal{H} = \mathcal{H}_S + \mathcal{H}_l + \mathcal{H}_{Sl}$, which comprises a spin $\mathcal{H}_S$, a lattice $\mathcal{H}_l$ , and a spin-lattice Hamiltonian, $\mathcal{H}_{Sl}$. The magnetic interactions, given by [16,17]

$$\mathcal{H}_S = \sum_{r=1}^{3} \frac{J_r}{2\hbar^2} \sum_{\langle ij\rangle_r} \boldsymbol{S}_i \cdot \boldsymbol{S}_j + \frac{1}{2\hbar^2} \sum_{\langle ij\rangle_2} \boldsymbol{D}_{ij} \cdot (\mathbf{S}_i \times \mathbf{S}_j) + \frac{K}{\hbar^2} \sum_i \left(S_i^z\right)^2 + \frac{g\mu_B B_z}{\hbar} \sum_i S_i^z$$

($S_i$ spin operator of site $i$, $\hbar$ Planck constant, g Landé factor, $\mu_B$ Bohr magneton) encompass Heisenberg exchange interactions up to third-nearest neighbors ($J_r$, r = 1, 2, 3), out-of-plane Dzyaloshinskii-Moriya interaction (DMI) between second-nearest neighbors ($\boldsymbol{D}_{ij} = \pm D\hat{\boldsymbol{z}}$) an easy axis anisotropy ($K$), and a Zeeman term that couples the spins to the external magnetic field $B_z$. For the lattice Hamiltonian [16]

$$\mathcal{H}_l = \sum_i \frac{(P_i^z)^2}{2M} + \frac{C}{2} \sum_{\langle ij\rangle} \left(u_i^z - u_j^z\right)^2$$

we focus on the out-of-plane vibrations as the in-plane phonon modes do not couple to the magnons to leading order (see below). Here, $M$ is the mass of the $Mn^{2+}$ ions, $\boldsymbol{P}_i^z$ is the $z$ component of the momentum operator of site $i$, and $\boldsymbol{u}_i^z$ is the displacement of site $i$ along $z$. The displacements are coupled to the spins by [16]

$$\mathcal{H}_{sl} = \frac{\lambda}{\hbar^2} \sum_{\langle ij\rangle 1} S_i^z \left(\boldsymbol{S}_i \cdot \hat{\boldsymbol{r}}_{ij}\right)\left(u_i^z - u_j^z\right)$$

where λ is the strength of the spin-lattice coupling and $\hat{\boldsymbol{r}}_{ij}$ is the normalized nearest-neighbor bond vector. We employ the following parameters specific to $MnPS_3$: $J_1$ = 1.054 meV, $J_2$ = 0.048 meV, $J_3$ = 0.3 meV, $D$ = 0.78 μeV, $K$ = -2 μeV, $g$ = 1.824, and λ = 0.0292 meV $Å^{-1}$ [16]. The spin quantum number of the local $Mn^{2+}$ spins is $S = 5/2$ **[53]**. For the elastic constant, we experimentally determined the sound velocity for the transverse phonons as 3238 m/s, which implies $C$ = 690.7 m $Å^{-2}$ assuming a lattice constant of $a$ = 5.88 Å.

*5.3.2 Mapping onto bosonic Hamiltonian.* In order to describe the spin degrees of freedom ($S_i$) by bosonic creation and annihilation operators

$(\boldsymbol{a_i}, \boldsymbol{a_i^\dagger})$, the truncated Holstein-Pimakoff transformation **[54]**

$$\frac{\boldsymbol{S_i}}{\hbar} = (\boldsymbol{S_i} - \boldsymbol{a_i^\dagger a_i})\boldsymbol{\hat{z}_i} + \sqrt{\boldsymbol{S_i}}(\boldsymbol{a_i^\dagger \hat{e}_i^+} + \boldsymbol{a_i \hat{e}_i^-})$$

is employed using the classical ground state spin orientations $\boldsymbol{\hat{z}_i}$, which play the role of local quantization axes. Here, $\boldsymbol{\hat{e}_i^\pm} = (\boldsymbol{\hat{x}_i} \pm \boldsymbol{i\hat{y}_i})/\sqrt{\boldsymbol{2}}$. The local axes $\boldsymbol{\hat{x}_i}, \boldsymbol{\hat{y}_i}$, **and** $\boldsymbol{\hat{z}_i}$ are chosen to constitute a right-hand coordinate system. Relabeling the sites *i* in terms of sublattice (*m*) and unit cell (*u*) indices, the Fourier transformation

$$\boldsymbol{a_{mu}} = \frac{\boldsymbol{1}}{\sqrt{\boldsymbol{N_{uc}}}} \sum_{\boldsymbol{k}} \boldsymbol{e^{ik\cdot(R_u+r_m)} a_{mk}}$$

$$\boldsymbol{a_{mu}^\dagger} = \frac{\boldsymbol{1}}{\sqrt{\boldsymbol{N_{uc}}}} \sum_{\boldsymbol{k}} \boldsymbol{e^{-ik\cdot(R_u+r_m)} a_{mk}^\dagger}$$

($N_{uc}$ number of unit cells, $R_u$ lattice vector of *u*-th unit cell, $r_m$ position of *m*-th basis site inside its unit cell) defines nonlocal bosonic modes.

For the vibrational degrees of freedom $(\boldsymbol{u_i^z}, \boldsymbol{p_i^z})$, we apply the transformation

$$\boldsymbol{u_i^z} = \sqrt{\frac{\hbar}{\boldsymbol{2M\omega}}}(\boldsymbol{b_i} + \boldsymbol{b_i^\dagger})$$

$$\boldsymbol{p_i^z} = \boldsymbol{i}\sqrt{\frac{\hbar\boldsymbol{M\omega}}{\boldsymbol{2}}}(\boldsymbol{b_i^\dagger} - \boldsymbol{b_i})$$

where $\boldsymbol{b_i}, \boldsymbol{b_i^\dagger}$ are bosonic annihilation and creation operators, and $\boldsymbol{\omega} = \sqrt{\frac{\boldsymbol{6C}}{\boldsymbol{M}}}$ is the local eigenfrequency.

*5.3.3 Intrinsic thermal Hall conductivity* The intrinsic contribution to thermal Hall conductivity is given by **[55,56]**

$$\boldsymbol{\kappa_{xy}} = -\frac{\boldsymbol{k_B^2 T}}{\hbar \boldsymbol{V}} \sum_{\boldsymbol{nk}} \boldsymbol{c_2[\rho(\epsilon_{nk})]\Omega_{nk}}$$

where $k_B$ is the Boltzmann constant and *V* is the system's volume. The sum runs over all bands n = 1, …, *N* (*N* is the number of bands) and all wave vectors *k*. The Berry curvature is given by **[57–59]**

$$\Omega_{nk} = i\hbar^2 \sum_{m=1,m\neq n}^{2N} \sum_{\mu,\nu=x,y} \epsilon_{\mu\nu} \frac{(Gv_{\mu,k})_{nm}(Gv_{\mu,k})_{mn}}{(\tilde{\varepsilon}_{nk} - \tilde{\varepsilon}_{mk})^2}$$

where $\epsilon_{\mu\nu}$ is the Levi-Civita symbol. The Berry curvature accounts for the anomalous transverse velocities of the magnon normal modes, while the distribution function $c_2(\rho) = (1+\rho)ln^2\left(\frac{1+\rho}{\rho}\right) - ln^2(\rho) - 2\text{Li}_2(-\rho)$, with $\rho = \rho(\epsilon_{nk}) = \left[e^{\beta\epsilon_{nk}} - 1\right]^{-1}$ being the Bose distribution function ($\beta = \frac{1}{k_B T}$) and $\text{Li}_2$ being the dilogarithm, accounts for the occupation number and energy carried by a quasiparticle.

## Acknowledgements

MN, JW, and JPH acknowledge funding support from the Center for Emergent Materials, an NSF MRSEC, under award number DMR-2011876. The work of AM was funded by the Deutsche Forschungsgemeinschaft (DFG, German Research Foundation) - Project No. 504261060 (Emmy Noether Programme).